# THERMAL TRANSIENT CHARACTEROZATION OF PACKAGED THIN FILM MICROCOOLERS

*Kazuhiko Fukutani, Rajeev Singh and Ali Shakouri*

Baskin School of Engineering, University of California Santa Cruz
1156 High Street, Santa Cruz, CA 95064, U.S.A.

**ABSTRACT**

A network identification by deconvolution (NID) method is applied to the thermal transient response of packaged and unpackaged microcoolers. A thin film resistor on top of the device is used as the heat source and the temperature sensor. The package and the bonding thermal resistances can be easily identified by comparing structure functions. High-speed coplanar probes are used to achieve a short time resolution of roughly 100ns in the transient temperature response. This is used to separate the thermal properties of the thin film from the substrate. The obtained thermal resistances of the buffer layer and Silicon substrate are consistent with the theoretical calculations. In order to estimate the superlattice thermal resistance and separate it from the thin $SiN_x$ layer deposited underneath the thin film resistive sensor, an order of magnitude faster thermal transient response is needed.

## 1. INTRODUCTION

Temperature control of microelectronic devices has become more important in recent years. Device miniaturization and higher switching speeds have increased power dissipation density substantially. Limiting the maximum die temperature has resulted in an increase of packaging complexity and cost. The failure rate due to electromigration and oxide breakdown, and the leakage power are exponentially dependent on temperature. Thus, thermal management both at the package and device levels plays a vital role in integrated circuit (IC) reliability and performance. Thin film silicon-based microcooler is one of the possible solutions for the thermal management of IC chips. This device has advantages over other cooling technologies as it enables monolithic integration for site-specific cooling for CMOS VLSI circuits. The silicon-germanium microcoolers have demonstrated 7 $^{o}$C cooling at 100 $^{o}$C and cooling power densities exceeding 500W/cm$^2$ [1]. These achievements are due to alloys and superlattice structures, which can increase Seebeck coefficient and decrease thermal conductivity without significantly reducing the electrical conductivity of the material [2]. By changing the alloy composition, superlattice period, thickness, doping concentration and device sizes, the cooling performance of microcoolers can be optimized [3]. However, when the microcoolers are mounted in a package or heat sink for measurements, the package and interface thermal resistances often affect the device's cooling performance significantly. Furthermore, as previously mentioned, overall cooling characteristics of the device strongly depend on the thermal properties of the thin film regions such as a superlattice and buffer layer. Therefore, accurate and simple techniques for measuring the thermal properties of different layers are critical for the design and optimization of microcoolers.

For a long time, thermal transient measurements have been used as one of the important methods for characterization of IC packages. These measurements can be used to separate different contributions to the total thermal resistance. They are also used to identify structure defects in the package. Temperature response measured by sensors for a given step excitation is widely used for analysis. In order to analyze the temperature response including structural information, various evaluation methods were proposed [4,5]. Among the evaluation methods, the technique proposed by Székely *et al*. [5], based on linear RC network theory, is very interesting because it can provide the map of heat current flow as a function of the cumulative resistance in the samples by using a new representation: the structure function. In their methods, the temperature response is transformed to the time constant spectrum of the investigated thermal model by deconvolution technique, and then, the time constant spectrum is transformed into the cumulative or differential structure function, defined as the cumulative thermal capacitance or the derivative of the cumulative thermal capacitance with respect to the cumulative thermal resistance [5]. By interpreting these functions, thermal resistances and capacitances of each part in the sample can be identified [5]. This is called network identification by deconvolution (NID) method.

Up to now, thermal transient analysis has been applied to various electronic and optoelectronic devices inside a package [5-9]. For example, the differences of partial and interfacial thermal resistance values among BJTs mounted to the heat sink by different methods were identified by using the differential structure function. Furthermore, for detecting voids in the die attach of packages, the use of differential structure function was successful. These results showed that NID technique is a powerful method to identify thermal resistances in the heat flow path.

However, there is no report about the thermal analysis of thin film devices by using this method. Since thermal time constant of micron thick thin films is an order of the microseconds or less and due to the small cross-sectional area of typical thin film devices, special sensors and high-speed measurement techniques are required. In this research, we tried to apply the NID method to thin film microcooler devices.

## 2. EXPERIMENT

The schematic structure of thin film Silicon based microcoolers is shown in Fig.1(a). In this research, SiGe/Si superlattice microcoolers were analyzed. The device consists of a 1µm thick Au contact electrode, 0.3µm $Si_{0.8}Ge_{0.2}$ cap layer with doping concentration of $1.9\times10^{20}$ cm$^{-3}$ (not shown), 3µm thick superlattice layer with the structure of 200×(3nm Si/12nm $Si_{0.75}Ge_{0.25}$), doped to $5\times10^{19}$ cm$^{-3}$, 1µm $Si_{0.8}Ge_{0.2}$ and 3µm graded SiGe buffer layer with the same doping concentration as the superlattice, and 500µm Si substrate.

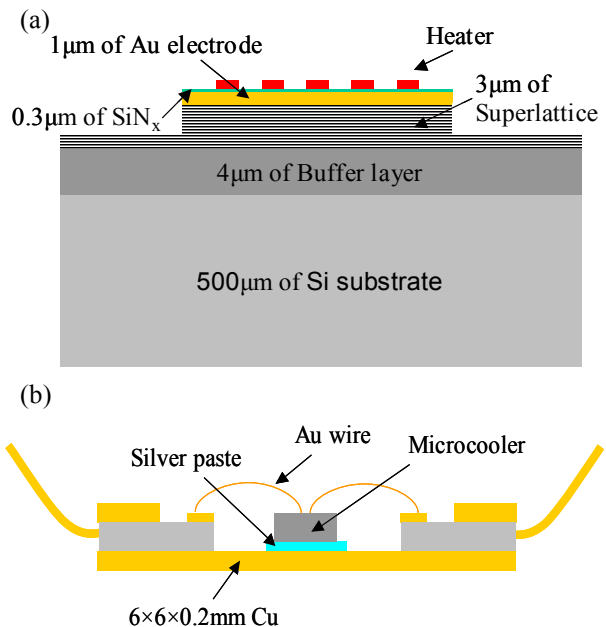

**Fig. 1.** (a)Schematic of the Si/SiGe superlattice microcooler and (b) packaged microcooler.

For a thermal transient characterization, heaters made by Au were deposited on the Au contact layer after a 0.3µm thick $SiN_x$ as electrically insulating layer. Some microcoolers are mounted on a package by a silver paste as adhesive materials, as shown in Fig.1 (b). All the microcoolers with and without the package are attached to a big Cu plate as the heat sink using thermal paste. Experiments were performed on superlattice microcoolers ranging in size from 70×70 to 100×100 microns square, with and without a package.

Figure 2 shows the SEM image of different size thin film SiGe/Si superlattice microcoolers fabricated using batch integrated circuit fabrication techniques. The heater structures on top of the devices can be clearly seen.

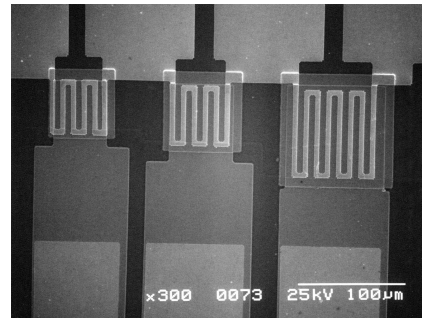

**Fig. 2.** SEM image of SiGe/Si superlattice microcoolers along with the thin film heaters/sensors.

Temperature response is measured using the heaters that are fabricated on top of microcoolers. The small heater can rapidly heat the sample. It can also be used as a temperature sensor. The temperature on the top of microcooler can be measured through the resistance change of the heater while the constant current as a step function is applied. The resistance of the heater, $R$, is expressed as $R(T) = R_0 + \alpha \times \Delta T$. $\alpha$ is the resistance change per Kelvin. By measuring sample's resistance at various ambient temperatures, $\alpha$ can be easily determined. The thermal transient measurement itself is very simple. A pulse current is applied to the heater on top of the microcooler through one set of probes. Simultaneously, the resistance's change of the heater is measured through other set of probes. The current and voltage signals from the heater are detected using a digital oscilloscope. When higher temporal resolution is necessary, a high-speed packaging and coplanar probes are used to reduce signal ringing due to electrical impedance mismatch. A short time resolution of roughly 100n sec. in thermal transient measurement has been achieved [10]. After measuring transient temperature response of the samples, an evaluation based on a distributed RC network theory was performed. Consequently, the time-constant spectrum and structure functions are calculated by direct mathematical transformation of the temperature response. Finally, the obtained experimental results were compared with the theoretical values, calculated using following material properties. In this table, $\kappa$ is the thermal conductivity and $c_v$ is the volume heat capacitance.

**Table 1** Thermal material properties for SiGe/Si microcooler [11,12]

| Material | $\kappa$ (W/mK) | $c_v$ (Ws/K·m$^3$) |
|---|---|---|
| PECVD SiN$_x$ | 0.7 | $4.25 \times 10^5$ |
| Superlattice | 8 | $1.67 \times 10^6$ |
| Buffer layer | 3 | $1.67 \times 10^6$ |
| Si sub. | 147 | $1.66 \times 10^6$ |

### 3. RESULTS AND DISCUSSIONS

Figure 3 shows the temperature transient responses of the Si/SiGe superlattice thin film microcoolers with and without a package at given step excitations in the 3μs – 10s range. The cross-sectional area of superlattice is 70x70 μm$^2$. These two chips are not completely identical. The size of the Si substrates is different. The cross-sectional areas of the Si substrate for the packaged and unpackaged samples are approximately 1.5×2.5 and 2.5×4 mm$^2$, respectively. Thin film superlattices for both samples are located near the edge of the die. The vertical axis denotes the temperature rise when heat power of 1W is given to the heater, which means that this axis also corresponds to the thermal resistance with the unit of K/W. Initial parts of the two curves almost coincide. This part should be related to the heat-flow inside the thin film superlattice and a part of the Si substrate. On the other hand, there is a big difference at the right side of the curves. This should be due to the package and additional interface materials, but it is difficult to interpret the detailed thermal effects of the package and the attachments from these curves.

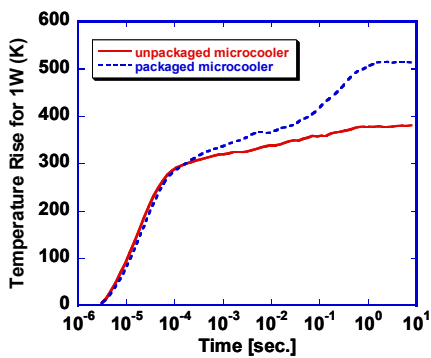

**Fig. 3.** Measured thermal transient curves of packaged and unpackaged microcoolers.

In order to understand the thermal properties of the package and additional interfaces, the NID method was applied to these heating curves by using direct mathematical transformation. After deconvolution operation using Bayes iteration [7], the time-constant spectrum of the heat-flow path could be obtained. Figure 4 shows time-constant spectra of each sample. The first large peaks at the time-constant on the order of 10$^{-5}$ sec. are almost the same, suggesting that these peaks are attributed to the thin film region and a part of substrate. Other intermediate small peaks are difficult to understand, but the large peak for the packaged sample at the time-constant of approximately 0.2-0.4 sec. is considered to be due to the package and additional interface materials.

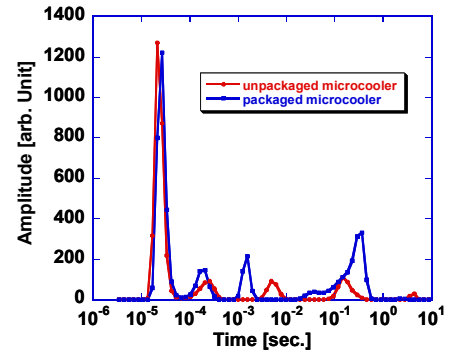

**Fig. 4.** Time constant spectra of packaged and unpackaged microcoolers.

Base on the obtained spectra, the cumulative structure functions were calculated and they are shown in Figure 5. The horizontal axis denotes the cumulative thermal resistance and the vertical one refers to the cumulative thermal capacitance. As one can see from the sample's structure with the package shown in Fig.1, the heat flow path is as follows: superlattice and buffer layer → Si substrate → silver paste → package → thermal paste → heat sink. The shape of the initial parts for both plots is very similar, suggesting that the left side of the arrows A and A' of Fig. 5 corresponds to a total thermal capacitance and resistance of the thin film region and Si substrate. The expected thermal capacitances for the Si substrate with a volume of 1.5×2.5×0.5 mm$^3$ (packaged sample) and 2.5×4.0×0.5 mm$^3$ (unpackaged sample) are calculated to be 0.0031 and 0.0083 Ws/K, respectively. These values are a little bit higher than the obtained cumulative thermal capacitances at the arrow A and A' of Fig. 5, which are approximately 0.001 and 0.005 Ws/K, respectively. The volume of the substrate for a heat flow path should be smaller than that of an actual substrate because of three-dimensional heat spreading and the location of the superlattice near the edge. Therefore, this result is quite reasonable. Furthermore, it makes sense that the sample mounted on the package has lower thermal capacitance and higher thermal resistance compared to the unpackaged sample. This suggests that the heat flow cross-sectional area for the packaged sample is smaller than that for the unpackaged one. These results are supported by the fact that the unpackaged sample consists of smaller Si substrate. The cumulative thermal resistances in the left side of the arrow A and A' (320~370 K/W) are related to the total thermal resistance of the thin film region and Si substrate. On the other hand, the right side of the arrow A' corresponds to the heat flow inside the package and the additional interface material because there is a big difference between the two plots. The thermal resistance of the package (Cu plate) is expected to be less than 1 K/W even if only the limited area that participates in the heat conduction is considered. Therefore, the first and second plateaus in the plot for the packaged sample are expected to be due to the attachment with the package (silver paste) and with the heat sink (thermal paste). From the plot, the thermal resistances of each attachment are estimated to be approximately 60 and 70 K/W respectively. As compared to the plot for the unpackaged sample, the thermal resistance of the attachment with heat sink for the packaged sample (~70 K/W) is quite larger than that for the unpackaged sample (~25 K/W), suggesting the presence of the poor thermal contact between the package and the heat sink. Since there is a possibility that the leads of the package could obstruct making a good thermal contact, it is necessary to pay attention to the contact with the heat sink when this kind of package is used. The presence of a void in thermal paste is also confirmed by the fact that the cumulative thermal capacitance is lower that that for unpackaged sample.

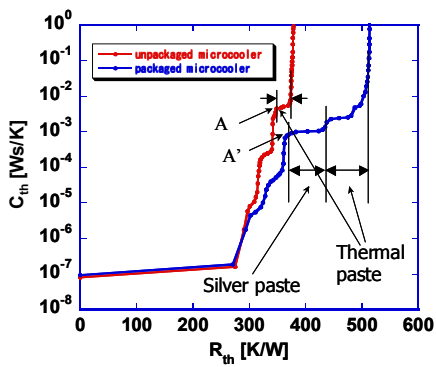

**Fig. 5.** Cumulative structure functions for packaged and unpackaged microcoolers.

As mentioned before, the fastest transient time that can be measured using a standard probe technique is on the order of $10^{-6}$ sec., due to electrical impedance mismatching. This speed is not high enough to distinguish the thermal resistance of the thin film region from the substrate because the time constant of thin film superlattice is typically on the order of $10^{-6}$ sec. Therefore, a high-speed measurement technique is used to reduce signal ringing. Figure 6 shows the temperature transient response for a Si/SiGe superlattice thin film microcooler when the high-speed measurement technique is used. In this sample, superlattice with a cross-sectional area of $100 \times 100$ μm$^2$ is located at a center of a substrate with a size of $2.5 \times 4 \times 0.5$ mm$^3$. The sample was not mounted on a package to avoid additional thermal resistances. One can clearly see a temperature response on the order of $10^{-7}$ sec without signal ringing. One may think that a higher temporal resolution is necessary to analyze thermal properties of thin film regions. However, as shown in Fig. 1(a), the insulation layer made by PECVD 0.3μm SiN$_x$ is deposited between top Au side contact electrode and heater. The thermal diffusion time through the SiN$_x$ film is typically on the order of $10^{-7}$ sec.. We think that the thermal diffusion time from $10^{-7}$ to 1 sec is due to the heat flow from a part of the SiN$_x$ through Au electrode, SiGe superlattice, buffer layer and Si substrate to the heat sink.

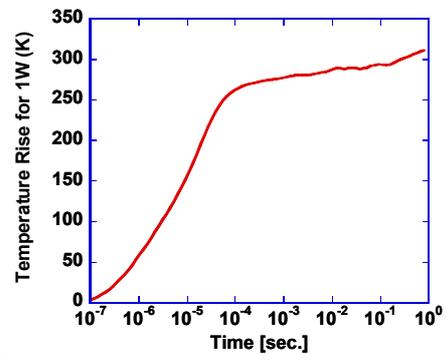

**Fig. 6.** Measured high-speed thermal transient of Si/SiGe superlattice thin film microoclers with a superlattice cross-sectional area of $100 \times 100$ μm$^2$.

Figure 7 shows the time constant spectrum of the sample. Two big peaks can be clearly seen at the short time scales from $10^{-7}$ to $10^{-4}$ sec. The peak at approximately $10^{-6}$ sec can't be seen in Fig .4. The central time constants of first and second big peaks are approximately $1.1 \times 10^{-6}$ and $1.6 \times 10^{-5}$ sec, respectively. The time constant of first peak is almost consistent with that estimated from thermal properties of 3μm thick SiGe superlattice. The second large peaks are considered to be due to the 4μm thick buffer layer and the Si substrate because these also match well with the estimated values. The small peaks on the order of $10^{-4}$ and $10^{-3}$ sec are unknown; they could be related to heat flow inside finite size Si substrate. There is another peak at approximately 0.3 sec, which is related to the interface between the microcooler and the heat sink. This peak can also be seen in the plot for the packaged sample, shown in Fig. 4.

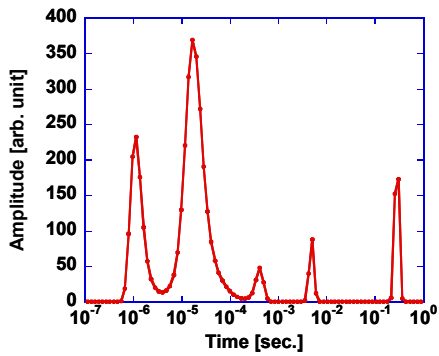

**Fig. 7.** Time constant spectrum of unpackaged SiGe/Si superlattice microcooler with high speed probes.

Figure 8(a) shows the cumulative structure function of the superlattice microcooler. The overall shape of the plot is very similar to that for unpackaged sample shown in Fig.5, but the detailed structure at initial heat flow path can be seen in this plot due to high-speed measurement. The thermal capacitance at the arrow A is approximately 0.009 Ws/K, which corresponds to the volume of $2.5 \times 4.0 \times 0.5$ mm$^3$ for the Si substrate. Therefore, the left and right sides of arrow A represent the heat flow inside the microcooler and outside it. The total thermal resistances of the microcooler and the interface between sample and heat sink are estimated to be approximately 289 and 25 K/W, respectively.

In order to analyze the detailed heat flow path inside the microcooler, a derivative of the cumulative structure function was calculated. Figure 8(b) shows differential structure function for the sample. The horizontal axis denotes the thermal resistance, the vertical one refers to the derivative of the cumulative thermal capacitance ($K$) with respect to the cumulative thermal resistance. The vertical axis, $K$, is equal to $c_v \cdot \kappa \cdot S^2$, where $c_v$ is the volume heat capacitance, $\kappa$ is the thermal conductivity and $S$ is the cross-sectional area in the heat flow path [7]. Based on the structure of the microcooler, the detailed heat flow path can be identified. The left side of the first arrow shown as A is due to heat flow inside superlattice, the curve between A and B is due to the buffer layer, the curve between B and C is due to Si substrate, and the curve between C and the end of the plot is due to the interface between microcooler and the heat sink. The value of $K$ at the arrow A is $2 \times 10^{-10}$ W$^2$s/K$^2$. Assuming that the material properties of SiGe superlattice are valid for this area, the cross-sectional area is calculated to be approximately $110 \times 110$ μm$^2$. Similarly, the value of $K$ at arrow B and C ($7.5 \times 10^{-8}$ and 0.024 W$^2$s/K$^2$) can be transformed to be the cross-sectional area of approximately $1.75 \times 10^{-2}$ ($132 \times 132$ μm$^2$) and 10.0 mm$^2$ ($2.5 \times 4$ mm$^2$), assuming the thermal conductivity and the volume heat capacitance for Si. The estimated cross-sectional area for SiGe/Si superlattice ($110 \times 110$ μm$^2$) is almost consistent with the actual area ($100 \times 100$ μm$^2$) within experimental errors. The cross-sectional area of $132 \times 132$ μm$^2$ for an initial heat spot on the Si substrate is larger than that of superlattice, but it can be explained by a three-dimensional heat spreading inside the buffer layer. On the other hand, the cross-sectional area of 10.0 mm$^2$ estimated for a heat exit on the backside of Si is completely consistent with the size of the Si substrate.

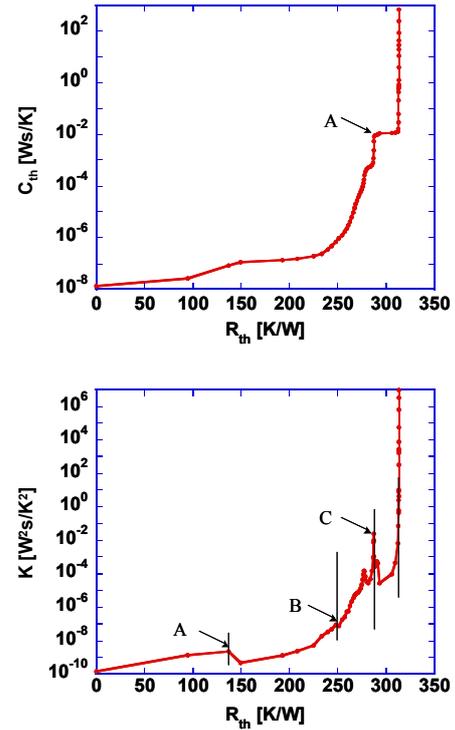

**Fig. 8.** (a) Cumulative structure function and (b) differential structure function for unpackaged SiGe/Si microcooler.

The obtained thermal resistances for different layers from this plot are summarized in Table.2.

**Table 2** Obtained thermal resistances for each layer

| Superlattice | Buffer layer | Silicon substrate | Interface |
|---|---|---|---|
| 136 K/W | 117 K/W | 36 K/W | 25 K/W |

The theoretical thermal resistance are shown in Table 3 when a one-dimensional heat transfer inside the superlattice and buffer layer region with $100 \times 100$ μm$^2$ cross-sectional area, and three-dimensional heat spreading inside the Si substrate with a heat source of $100 \times 100$ μm$^2$ cross-sectional area are assumed [13].

**Table 3** Theoretically calculated thermal resistances for each layer

| Superlattice | Buffer layer | Si substrate |
|---|---|---|
| 37.5 K/W | 133 K/W | 32.3 K/W |

Comparing Tables 2 and 3, it is clear that the calculated thermal resistances are quite similar except for that of the superlattice. Why is the thermal resistance of the superlattice so different? This may be explained by the speculation that the heat from a thin film resistor is localized inside $SiN_x$ since an effective area of the heater is $61.8 \times 61.8$ μ$m^2$, as shown in Fig. 2. The time-constant of 0.3 μm $SiN_x$ is estimated to be on the order of $10^{-7}$ sec due to localized heating, which indicates that an initial temperature rise on the order of $10^{-7}$ sec is strongly affected by the thermal resistance of $SiN_x$. Furthermore, the localized heating provides large thermal resistance of 112 K/W for 0.3 μm $SiN_x$, calculated from an effective area of the heater ($61.8 \times 61.8$ μ$m^2$). If the obtained thermal resistance for superlattice is assumed to be including that for $SiN_x$, the thermal resistance for superlattice is quite reasonable. The problem for our evaluation is as follows. In our mathematical transformation from a transient curve to time constant spectrum, we assume that the temperature rise in less than $10^{-7}$ sec is zero. This is not true for real samples because heat flows inside $SiN_x$ in this time scale, and this enhances the height of the peak for superlattice in time constant spectrum to compensate the thermal resistance of the $SiN_x$. Therefore, in order to separate the thermal resistances of $SiN_x$ and the superlattice, a shorter time resolution on the order of $10^{-8}$ sec is necessary.

## 4. CONCLUSIONS

The thermal response curves of packaged and unpackaged microcoolers were analyzed by NID method based on linear RC network theory. A good thermal contact between the sample and the heat sink can give a thermal resistance of approximately 25 K/W. However, a bad contact including voids induces additional thermal resistances. A high-speed measurement using coplanar probes and high-speed packaging could achieve a short time resolution of roughly 100ns in temperature response curve. This is used to separate the thermal resistances of the buffer layer and the substrate. The obtained thermal resistances of buffer layer and Si substrate are 117 and 36 K/W, which are almost consistent with the theoretical values. Therefore, the NID method is very useful for the thermal evaluation of thin film devices if we can achieve very short time resolution.

## 5. ACKNOWLEGEMENT

The authors would like to acknowledge helpful discussions with Dr. Yan Zhang (Flomerics Inc.). This work was supported by Canon Inc. and Office of Naval Research.